\documentclass[12pt]{article}

\def\intB_#1^#2{({\bf{B}})\mspace{-7mu}\int_{#1}^{#2} }
\def\intBR_#1{({\bf{B}})\mspace{-7mu}\int_{#1}}

\def \BA   {\left[\begin{array}}
\def \EA   {\end{array}\right]}
\def \BEQ  {\begin{equation}}
\def \EEQ  {\end{equation}}

\def\eqbydef{\mathrel{\stackrel{\Delta}{=}}}

\def\eqbydef{\mathrel{\stackrel{\Delta}{=}}}

\def \Rset{\mathbb{R}}

\newcommand{\finp}{\raisebox{-0.0ex}{\rule{1.4ex}{1.4ex}}}

\usepackage{graphics} 
\usepackage{epsfig} 
\usepackage{mathptmx} 
\usepackage{times} 
\usepackage{amsmath} 
\usepackage{amssymb}  

\newtheorem{problem}{Problem}

\newtheorem{proposition}{Proposition}

\newcommand\reals{\mathbb{R}}
\def\eqbydef{\mathrel{\stackrel{\Delta}{=}}}

\usepackage[backend=biber,url=false,doi=false,isbn=false,eprint=false]{biblatex}
\bibliography{../../../bibFiles/dinhBio.bib, ../../../bibFiles/dinh.bib}

\usepackage{color}

\begin{document}

\title{A RBA  model for the chemostat modeling}
\author{Marc Dinh and Vincent Fromion\footnote{M. Dinh and V. Fromion are with MaIAGE, INRA, Universit\'{e} Paris-Saclay, 78350, Jouy-en-Josas, France,
        {\tt\small marc.dinh@inra.fr}
and
        {\tt\small vincent.fromion@inra.fr}}}
\date{\today}
\maketitle

\begin{abstract}
The purpose of this paper is to show that it is possible to replace Monod's type model of a chemostat by  a constraint based model of bacteria at the genome scale. This new model is an extension of the RBA model of bacteria developed in a batch mode to the chemostat.  This new model, and the associated framework, leads to a dramatic improvement in the prediction capacities of the chemostat behaviour. Indeed, for example, the internal states of the bacteria are now part of the prediction outputs and the chemostat behaviour can now be predicted for any limiting source. Finally, the first interests of this new predictive method are illustrated on a set of classic situations where predictions are already close of the well-known biological observations about chemostat.

This paper is an extended version of \cite{DiF:19} that includes a discussion on the modeling assumptions.
\end{abstract}


\section{Introduction}
The chemostat is a specific and popular experimental method introduced
in the field of microbiology in the early
50s in \cite{Mon:50,NoS:50,NoS:50a}. The chemostat consists in growing
micro-organisms in a  culture vessel where a defined fresh medium is pumped from a reservoir into the culture vessel at a given flow rate (called the inflow or dilution  rate). As the culture liquid is removed at the same flow rate than the inflow, the volume of culture remains  constant.
By providing essential substrates / nutrients in excess at the exception of a limiting one, the chemostat generally ensures the
perpetual regeneration of the micro-organisms at a growth rate controlled by the limiting nutrient (remember that a part of the micro-organisms is removed by the outflow). This growth rate is equal to the dilution rate.

Experimental data have validated this general principle and shown that a chemostat inoculated with an unique micro-organism reaches, after a transient phase, a steady-state regimen where the micro-organism population and all the substrates concentrations in the culture vessel are constant. This regimen, called a balanced regimen, is used to study the physiology of micro-organism(s) under stable and controlled  conditions and for a wide range of flow rates and thus of growth rates. For example, the chemostat is used to investigate if the micro-organism population uses energy for maintenance  (at the growth rate close to $0$) and if, more generally, the biomass production yield with respect to the limiting energy source is constant or not, see e.g. \cite{MNC:63,Pir:65,Pir:72,StB:73,RuB:79,TeN:84,VSS:91,SLS:94}.

The chemostat is also useful for the experimental study of the evolution of living organisms, including the evolution
and competition between genetic variants within a population (such an issue is already considered in \cite{NoS:50a}) or even between populations of different  micro-organism species, see  e.g. \cite{StL:73,HHW:77,Vel:77} and also the textbooks \cite{Wal:83,SmW:95}. Finally, the chemostat is widely used in biotechnology  as a mean to continuously produce molecules of interest such as vitamins, see e.g. \cite{DaS:01,DSH:02} and more generally \cite{Doc:08}.

Another specific feature of the development of the chemostat is that it has been accompanied since its origin by the development of dedicated mathematical models describing how they work, see \cite{Mon:50,NoS:50a,HET:56}. The phenomenological model developed in \cite{Mon:50} is nowadays known as Monod's model of the chemostat.  More importantly from our point of view, the chemostat has stimulated interactions between the modeler / mathematical and biological communities, which have led to significant progress on fundamental issues such as those related to the evolution and competition between populations of micro-organism species (including issues related to microbial ecology).

However, the phenomenological nature of Monod-type models implies that some essential questions are currently beyond any theoretical investigation. In fact, to consider them, it is clearly necessary to have chemostat models functioning with extended predictive capabilities. This clearly requires a finer modeling of the micro-organism and in particular to have internal / infra-scale models of cells, also called whole cell models in the literature, that would somehow and ideally predict the behaviour of cell regardless of the composition of the culture medium. This last claim is the origin of this work, since our aim is to investigate if and how the new constraint-based method called {\em Resource Balance Analysis} (RBA) introduced in \cite{GFS:09, GFS:11}, which performs quantitative predictions of the internal behavior of
cells \cite{GMC:15},  can be extended to the chemostat.

We first recall that at the center of the RBA method is the concept of
resource allocation between the cellular processes, see \cite{GoF:17} for a review. RBA formalizes the relationships defining the interactions and allocation of resources between the cellular processes in a balanced regimen as a set of linear constraints. Actually,  RBA predicts, for a given medium defined by a vector of substrate / nutrient concentrations, the (possibly empty) convex set of all feasible phenotypes that ensures either cell viability only (no growth) or growth at a given growth rate. For a given growth rate,
each point of this set (when non empty) corresponds to a bacterial cell
configuration compatible with the chosen growth rate. The bacterial cell
configuration is mainly defined by the abundance of each molecular
machine present within the cell, i.e. the enzymes catalyzing the
metabolic reactions or the ribosomes producing the proteins to cite the
two main sets of machines, and by the activity of all molecular
machines, such as the metabolic fluxes associated for the enzymes or the flow of protein production for the ribosomes.

Like other constrained-based methods, see e.g. \cite{VaP:94,BBLE:17}, the RBA method calls the resolution of linear programming problems, and from
this point of view, fulfills one of the central issues of the
development of predictive methods for internal models of cells, i.e.  the capability to compute
predictions efficiently despite the large model size. The prediction of the cell configuration for a large numùber of medium configurations, and thus the prediction of cell regulations such as the well-known catabolite repression or lesser known ones \cite{TGF:17}, is then possible.

\medskip

This article is another investigation on the potentialities of the RBA method which is now made easier to the community by the software RBApy \cite{BFD:19}. Its main objective is to show that the method, introduced in batch mode in \cite{GFS:09, GFS:11} and experimentally validated in \cite{GMC:15}, can be extended to the chemostat problem. The model being presented in Section \ref{sec:model}, it is shown in Section \ref{sec:solution} that the problem of determining the balanced regimen of the chemostat can be obtained by solving by dichotomy (on the population variable) a series of convex optimization problems, under realistic and mild assumptions on the modeling of transporters. In order to show the interests of this new method, we compare in Section \ref{sec:validation} 
a first set of predictions with results available in the biological literature, and show the remarkable agreement with the known behaviour of bacteria in chemostat. Section \ref{sec:conclusion} concludes the paper.

\medskip

{\em Notation.} The symbol $\reals$,  $\reals_\geq$ and $\reals_>$ denote the set of real,  non-negative real and positive real numbers respectively. $\reals^n$ and $\reals^{m\times n}$ denote the set of real vectors of length $n$ and real matrices of size $m \times n$ respectively. $X^\intercal$ denotes the transpose of $X$. For a vector $x\in \Rset^n$, $x_i$ denotes its $i$-th component. Finally, the symbol $\eqbydef$ means ``equal by definition''.

\section{RBA Model for the bacteria in a chemostat}\label{sec:model}

\subsection{The chemostat model}
When the chemostat is well-stirred, its dynamical model can be described (see e.g. \cite{BaD:90,Doc:08} for details)  by the following system of differential equations:
\begin{equation}
\left\{\begin{array}{rcl}
dX(t)/dt & = & \left[\mu(S(t)) - D \right]X(t) \\[4pt]
dS_i(t)/dt & = & D \left[\bar{S}_i - S_i(t) \right] -  X(t) \kappa_i(S(t)), \ \text{$i\in\mathcal{S}$}
\end{array}
\right.
\label{full_chemostat_model}
\end{equation}
where\footnote{The units are: $g_{DW}$ for gram of cell dry weight, $L$ for liter, $h$ for hour and $mmol$ for millimole.}
\begin{itemize}
\item $X(t) \in\reals_\geq$ denotes the concentration of the cell population in the chemostat ($g_{DW}/L$);
\item $D \in\reals_\geq$ denotes the dilution rate of the chemostat, more precisely it is the percentage of the chemostat volume that is replaced per unit of time ($h^{-1}$);
\item $S(t)\in\reals_\geq^{n_s}$ denotes the vector of concentrations of the $n_s$ substrates in the chemostat medium ($mmol/L$). The set of indices of substrates is denoted by ${\cal S}$: ${\cal S} \eqbydef  \{1,\dots,n_s\}$;
\item $\bar{S} \in\reals_\geq^{n_s}$ denotes the vector of concentrations of the $n_s$ substrates in the chemostat inflow ($mmol/L$);
\end{itemize}
and where  
\begin{itemize}
\item $\mu$ is a function defined from $\reals_\geq^{n_s}$ into $\reals_\geq$ which provides the value of the growth rate (in $h^{-1}$) of the population for a given vector of substrate concentrations $S$; 
\item $\kappa$ is a function defined form $\reals_\geq^{n_s}$ into $\reals^{n_s}$ which provides the vector of exchange fluxes per unit of cell population ($mmol/g_{DW}/h$). By convention, the $i$-th  component of the vector $\kappa$ is non-negative when the cell imports the $i$-th substrate.
\end{itemize}

\medskip

We associate to the chemostat model this first problem.
\begin{problem} \label{pb:chemostat}
For given $D>0$, $X>0$ and vector of inflow concentrations $\bar{S}\in\reals_\geq^{n_s}$, find if there exists a vector $S^* \in \reals_\geq^{n_s}$ such that the system of equalities
\begin{displaymath}
\left\{ \begin{array}{l}
\mu(S^*) = D \\
D(\bar{S}_i -  S^*_i) = X\kappa_i(S^*),\ \text{ for all } i \in {\cal S} \end{array} \right.
\end{displaymath}
holds. 
\end{problem}

\subsection{Monod's type chemostat model}
Monod's chemostat model  derived in \cite{Mon:50} corresponds to experimental set-ups where the concentration of one substrate / nutrient is chosen to be limiting and the other ones are chosen to be largely in excess or are being controlled \cite{SHH:96,DaS:01,DSS:01,DSH:02}. That led J. Monod to develop a model where only the limiting substrate is considered, i.e. to the case $n_s\eqbydef 1$. 

Furthermore, the experimental data indicates that there exists an empirical relationship between the growth rate function $\mu$  and the uptake function $\kappa$ when they are considered as functions of the limiting substrate, i.e., 
$$
\begin{array}{ccc}
\displaystyle \mu(S) \eqbydef  \frac{\mu_{max} S}{K_S + S}  & \text{and} & \displaystyle \kappa(S) \eqbydef  \frac{1}{Y_S} \frac{\mu_{max} S}{K_S + S} 
\end{array}
$$
where $\mu_{max}$ is the maximum specific growth rate, $K_S$ is the saturation constant and $Y_{S}$ is a coefficient defined by Monod as an efficiency constant, which is nowadays referred to the biomass production yield with respect to the limiting resource $S$.
It is also shown in \cite{Mon:50} (see also \cite{HET:56})  that the obtained chemostat model has an unique stable equilibrium point ($X^*,S^*)^\intercal$ when $D <\mu_{max}$ (and implicitly also when $D$ is greater to some minimal value greater than 0) and that this equilibrium is such that $X^* \eqbydef Y_S (\bar{S}-S^*)$ and  $\mu(S^*)\eqbydef D$. The latter relation is well-known and indicates that, in a balanced regimen, the micro-organism population in the chemostat has a growth rate equal to the dilution rate. 

Finally, straightforward computations lead to the explicit expressions of the equilibrium components:
$$
\begin{array}{rcl}
S^*&\eqbydef & \displaystyle K_S \frac{D}{\mu_{max}-D}\\[6pt]
X^* &\eqbydef& Y_S (\bar{S}-S^*) = \displaystyle Y_S\left(\bar{S}-K_S \frac{D}{\mu_{max}-D}\right).
\end{array}
$$


Monod's model, represented by the functions $\mu$ and $\kappa$, is unable to predict the internal behaviour of a cell, which is one of the objective of this article. The purpose of the next section is to present a detailed way, namely the RBA method, to represent this behaviour, allowing a large set of substrates as well as complex behaviors.

\subsection{The RBA  model of bacteria in the chemostat context} \label{Assumption_convex_simplication}
The problem described below is obtained with the following assumptions: \emph{for each substrate} in the medium,
\begin{enumerate}
\item it is either imported or exported but not both. The set of imported substrates is denoted $\mathcal{S}_i$ while the set of exported ones is denoted $\mathcal{S}_e$;
\item there is one transporter in the cell;
\item if it is imported, the transporter efficiency is described by a Michaelis-Menten function;
\item if it is exported, the transporter efficiency is a constant.
\end{enumerate}
These assumptions are actually realistic as shown by the validation results of Section \ref{sec:validation} and as argued in the appendices \ref{sec:discuss TransporterEfficiency} and \ref{sec:one_transporter}. The RBA problem can then be formulated as the following linear programming problem (see Appendix \ref{sec:rbaDesc} or \cite{BFD:19} for a short description and \cite{GFS:09, GFS:11,GMC:15} for a detailed description).
\begin{problem} \label{pb:RBA}
For given $\mu \geq 0$ and $S\in\reals_\geq^{n_s}$, find $E\in\reals_\geq^{n_E}$, $T\in\reals_\geq^{n_T}$, $R\geq0$, $C\geq 0$, $\eta\in\reals^{n_E}$ and $\nu\in\reals^{n_T}$ such that the following constraints hold:
\begin{description}
\item[$(C_1)$] Mass conservation:
  \begin{description}
  \item[$a)$] $\mu\left( C_E^pE + C_T^pT + C_R^pR + C_C^pC + C_G^pP_G \right) - S^p v_f = 0$,
  \item[$b)$] $\mu\bar{X}^c - S^cv_f = 0$,
  \item[$c)$] $\mu\left( C_E^rE + C_T^rT+ C_R^rR + C_C^rC + C_G^rP_G \right) + S^rv_f =0$,
  \item[$d)$] $S^iv_f =0$;
  \end{description}
\item[$(C_2)$] Translation apparatus and chaperon folding capacity:
  \begin{description}
  \item[$a)$] $\mu\left( C_E^RE + C_T^RT+ C_R^RR + C_C^RC + C_G^RP_G \right) \leq k_TR$,
  \item[$b)$] $\alpha_c\mu\left( C_E^RE + C_T^RT+ C_R^RR + C_C^RC + C_G^RP_G \right) \leq k_CC$;
  \end{description}
\item[$(C_3)$] Cytoplasm (density) and membrane (surface) occupancy:
  \begin{description}
  \item[$a)$] $C_E^{D\intercal}E + C_R^{D}R + C_C^{D}C + C_G^DP_G \leq D^D$,
  \item[$b)$] $C_T^{S\intercal}T+ C_R^{S}R + C_C^{S}C + C_G^{S}P_G \leq D^S$;
  \end{description}
\item[$(C_4)$] (Internal) enzymatic capacity: $-\underline{k}_iE_i \leq \eta_i \leq \overline{k}_iE_i$, for all $i\in{\cal E}$;
\item[$(C_5)$] Transporters capacity:
  \begin{description}
  \item[$a)$] $0 \leq \nu_i \leq \displaystyle{\frac{V_iS_i}{K_i+S_i}}T_i$, for all $i\in{\cal S}_i$, with $V_i > 0$ and $K_i > 0$,\\
  \item[$b)$] $0 \leq -\nu_i \leq V_iT_i$, for all $i\in{\cal S}_e$, with $V_i > 0$;
\end{description}
\end{description}
where $v_f^\intercal \eqbydef \begin{pmatrix} \nu^\intercal,  \eta^\intercal \end{pmatrix}$ and where $\mathcal{E}$ denotes the set of internal enzymes, that is the set of enzymes that are not transporters.
\end{problem}

\medskip

The decision variables $E$, $T$, $R$ and $C$ are, respectively, the vectors of concentrations of internal enzymes, transporters, ribosome and chaperon ($mmol/g_{DW}$); $\nu$ and $\eta$ are, respectively, transporter and internal fluxes per unit of cell population ($mmol/g_{DW}/h$). For the purpose of this paper, we stress that the constraint $(C_{5-a})$ is a finer description of the constraint $(C_4)$ for the import transporters. Needed for the proofs, we also notice  that all the data in $(C_3)$ are by definition non-negative, meaning more precisely that the entities actually take some place in the cytosol or the membrane.

\medskip

For notational convenience, we introduce the variable $P^\intercal \eqbydef \begin{pmatrix} T^\intercal,E^\intercal, R,C \end{pmatrix}$ and $Lin_{RBA}[\mu](P,v_f)$ which consists in the constraints from $(C_1)$ to $(C_4)$ and $(C_{5-b})$. Problem \ref{pb:RBA} then reads: for given $\mu \geq 0$ and $S\in\reals_\geq^{n_s}$, find $P\in\reals_\geq^{n_E+n_T+2}$ and $v_f\in\reals^{n_E+n_T}$ such that $Lin_{RBA}[\mu](P,v_f)$ and $(C_{5-a})$ hold.

\subsection{A RBA model in a chemostat}
We then reformulate Problem \ref{pb:chemostat} in the RBA framework.
\begin{problem} \label{pb:chemostatRBA}
For given $D>0$, $X>0$ and vector of inflow concentrations $\bar{S}\in\reals_\geq^{n_s}$,  find if there exist vectors $S^*\in\reals_\geq^{n_s}$, $P^*\in\reals_\geq^{n_E+n_T+2}$ and $v_f^*\in\reals^{n_E+n_T}$ 
such that the following constraints:
\begin{equation} \label{eqn:basicConditionsGen}
\left\{ \begin{array}{l}
D(\bar{S}_i -  S_i^*) = X \nu_{i}^*,   \text{ for all } i\in{\cal S} \\[3pt]
 \displaystyle 0 \leq D(\bar{S}_i -  S_i^*) \leq X\frac{V_i S_i^*}{K_i+S_i^*}T_i^*, \text{ for all } i\in{\cal S}_i \\[6pt]
Lin_{RBA}[D](P^*,v_f^*)
\end{array} \right. 
\end{equation}
hold.
\end{problem}

\section{Resolution of the chemostat problem} \label{sec:solution}

In this section, we prove that Problem \ref{pb:chemostatRBA} is a convex problem.  
We first prove that the resolution of Problem \ref{pb:chemostatRBA} can include only the import transporters where $\bar{S}_i > 0$ with $i\in{\cal S}_i$.
\begin{proposition} \label{prop:simplifyPb}
Assume that Problem \ref{pb:chemostatRBA} is feasible and let $(S^*,P^*,v_f^*)$ be a solution of Problem \ref{pb:chemostatRBA}. Then  for every $i\in{\cal S}_i$, we have 
\begin{displaymath}
\left\{ 
\begin{array}{l}
\bar{S}_i=0 \ \Rightarrow \ S_i^* = 0 \\[2pt]
\bar{S}_i>0 \ \Rightarrow \ S_i^* \in(0,\bar{S}_i].\\
\end{array} 
\right.
\end{displaymath}
\end{proposition}
{\em Proof.} Let us assume that $\bar{S}_i\eqbydef 0$ for an $i\in{\cal S}_i$. We have $0 \leq -DS_i^*$. But since $S_i^*$ is non negative and $D > 0$, $S_i^*$ is necessarily null. The proof of $S_i^* > 0$ is performed by contradiction: assume $\bar{S}_i > 0$ and $S_i^* = 0$ for some $i\in{\cal S}_i$. It  then implies that $0 \leq D\bar{S}_i \leq 0$, i.e. $\bar{S}_i = 0$ since $D>0$, which is a contradiction. Finally, $S_i^* \leq \bar{S}_i$ comes from $0\leq D(\bar{S}_i-S_i^*)$ together with $D > 0$. \hfill \finp

\medskip

We can then now present the main result of the paper.
\begin{proposition} \label{prop:basicConvex}
Problem \ref{pb:chemostatRBA} is a convex feasibility problem.
\end{proposition}
{\em Proof.} By virtue of Proposition \ref{prop:simplifyPb}, only the transporters associated to non-zero $\bar{S}_i$ with $i \in{\cal S}_i$ have to be considered. Looking to the remaining constraints, only the constraints 
\begin{displaymath}
D(\bar{S}_i -  S_i)\leq X\frac{V_i S_i}{K_i+S_i}T_i,\text{ for all } i \in{{\cal S}_i} \text{ such that }\bar{S}_i>0 
\end{displaymath}
are potentially non-convex, being the only ones which are nonlinear. Thus if these constraints are actually convex, the overall problem is convex. The result follows from the fact that in this case,  the decision variable $S_i$ associated to an $i\in {\cal S}_i$ such that $\bar{S}_i>0$ belong to $(0,+\infty)$ which is a convex set.
For every $S_i \in (0,+\infty)$, the previous constraint can be rewritten as 
\begin{displaymath}
D \left( \frac{\bar{S}_i K_i}{S_i} - S_i + (\bar{S}_i- K_i) \right) - X V_i T_i \leq 0
\end{displaymath} 
which is a convex constraint as the left-hand side function is convex being the sum of convex functions.
\hfill \finp

\medskip

Actually, it is interesting to characterize the set of $X$ values such that Problem~\ref{pb:chemostatRBA} remains feasible. To do so, we introduce the following optimization problem.
\begin{problem} \label{pb:chemostatRBAMax}
For a given $D>0$ and a vector of inflow concentrations $\bar{S}\in\reals_\geq^{n_s}$, the chemostat with RBA maximization problem is defined by 
\begin{displaymath}
\displaystyle\underset{X\in\reals_\geq,\,  S^*\in\reals_\geq^{n_s}, \,  P^*\in\reals_\geq^{n_E+n_T+2}, \, v_f^*\in\reals^{n_E+n_T}} {\sup \ \ X}
\end{displaymath}
such that the constraints \eqref{eqn:basicConditionsGen} hold.
\end{problem}

\medskip

The next  result indicates that this problem has a maximum and that this maximum can be obtained by performing a dichotomy on $X$. 
\begin{proposition} \label{prop:dichotomy}
Assume that Problem \ref{pb:chemostatRBAMax} is feasible and let $X^*$ denote the supremum. 
Then the following statement holds:
\begin{enumerate}
\item[$(i)$] the supremum of Problem \ref{pb:chemostatRBAMax} is a maximum;
\item[$(ii)$] for every $X$ such that $0 \leq X \leq X^*$, Problem \ref{pb:chemostatRBA} is feasible.
\end{enumerate} 
\end{proposition}
{\em Proof.} 
{\em Let us first prove $(i)$.}  It relies on the extreme value theorem. Since the objective function is continuous, the result is obtained if the feasible set is compact, that is bounded and closed. The closedness of the feasible set is proved using the facts that the intersection of a finite number of closed sets is closed and that a set defined as the pre-image of a closed set by a continuous function is closed.

The boundedness is proved as follows. Since we have assumed that all entities actually take some place somewhere, the entities are bounded by the constraint $(C_3)$, that is $P$ is bounded. Thus the fluxes $v_f$ are also bounded by the constraints $(C_4)$ and $(C_5)$. For the constraint $(C_5-a)$, we need to notice that $S_i^*\in(0,\bar{S}_i]$, $i\in\mathcal{S}_i$, so that the efficiency of the transporter is also bounded. And since $S_i^*\in(0,\bar{S}_i]$, $i\in\mathcal{S}_i$, these $S_i^*$ are bounded; $X$ is  also bounded due to the constraints $D(\bar{S}_i -  S_i^*) = X\nu_{i}^*$ for all $i \in {\cal S}_i$. Finally, $S_i^*$, $i\in\mathcal{S}_e$, is also bounded due to same constraint.

%
%

\medskip

{\em Let us prove $(ii)$.} Let us denote $(X^*,S^*,P^*,v_f^*)$ a solution of Problem \ref{pb:chemostatRBAMax}. Let $X$ be such that $0 \leq X \leq X^*$. We construct $(S,P,v_f)$ to be a solution of Problem \ref{pb:chemostatRBA}. We set $P = P^*$, $v_f = v_f^*$ so that $Lin_{RBA}[D](P,v_f)$ holds. Finally we set $S$ such that $D(\bar{S}_i - S_i) = X\nu_{i}, \ i\in{\cal S}$. We show that such $S_i$ are non-negative. For each $i\in\mathcal{S}_e$, we have $\bar{S}_i - S_i \leq 0$ since $\nu_i \leq 0$, that is $S_i \geq \bar{S}_i \geq 0$. Since $X \leq X^*$, for each $i\in\mathcal{S}_i$, we have $\bar{S}_i-S_i \leq \bar{S}_i-S_i^*$, that is $S_i \geq S_i^* \geq 0$. Finally, since we have
\begin{displaymath}
0 \leq \frac{V_i S_i^*}{K_i+S_i^*} \leq \frac{V_i S_i}{K_i+S_i},\ \text{ for all } i\in{\cal S}_i,
\end{displaymath}
$(C_{5-a})$ holds since $T_i$ is non-negative, which completes the proof. \hfill \finp

\medskip

\section{A first validation of RBA prediction} \label{sec:validation}
Since we have already shown in \cite{GMC:15} that the predictions made by the RBA method are consistent with the experimental data when the concentration of substrates in the culture medium is known, the purpose of this section is to illustrate the potential of the RBA approach to capture some well established behaviours observed through biological experiences  with  chemostats. All illustrations are then made by adapting to the chemostat problem the RBA model of {\it Bacillus subtilis} developed and calibrated in \cite{GMC:15}.  Finally, we note that the RBA predictions reported in the following are those that correspond to the resolution of Problem \ref{pb:chemostatRBAMax} where the bacteria population is maximized.

\subsection{Chemostat in glucose-limited conditions}
\begin{figure}[thpb]
\centering
\includegraphics[width=12cm]{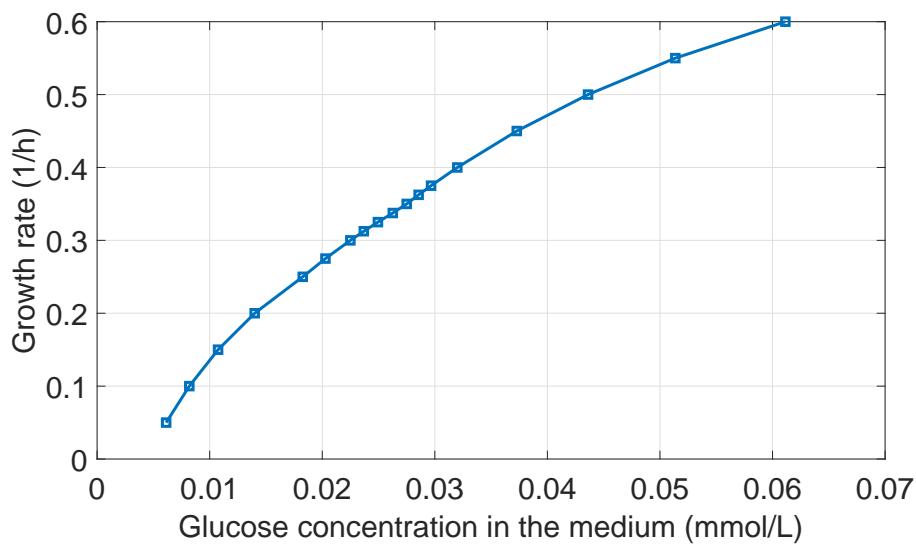}
\caption{``Monod's  curve'' in glucose-limited conditions} \label{fig:glc_limited_monod}
\end{figure}

We consider the experimental conditions defined in \cite{DaS:01} where the limited nutrient is the glucose. Following \cite{DaS:01}, we assume that the input concentration of glucose is $34$  $mmol/L$.  Our first question is whether we obtain, when we change the chemostat dilution rate, the classic characteristic of the growth rate as a function of the concentration of the limiting substrate. This curve is similar to the one that was adjusted by Monod to develop his model. In fact, the resulting curve described in Fig.~\ref{fig:glc_limited_monod} has the expected characteristics, except that a positive offset concentration in the medium is needed even at a null growth rate. This offset results from the cell's use of energy in order to maintain constant certain physiological quantities such as e.g. the osmotic pressure or the pH of the cytosol. The initial Monod model was modified to integrate a maintenance cost \cite{Pir:65} but its existence was debated during a long time due to the difficulty of measuring this maintenance. Fig. \ref{fig:glc_limited_monod} is actually a Pirt curve (see also comments on carbon distribution of Fig. \ref{fig:glc_limited_end_product}).

\begin{figure}[thpb]
\centering
\includegraphics[width=12cm]{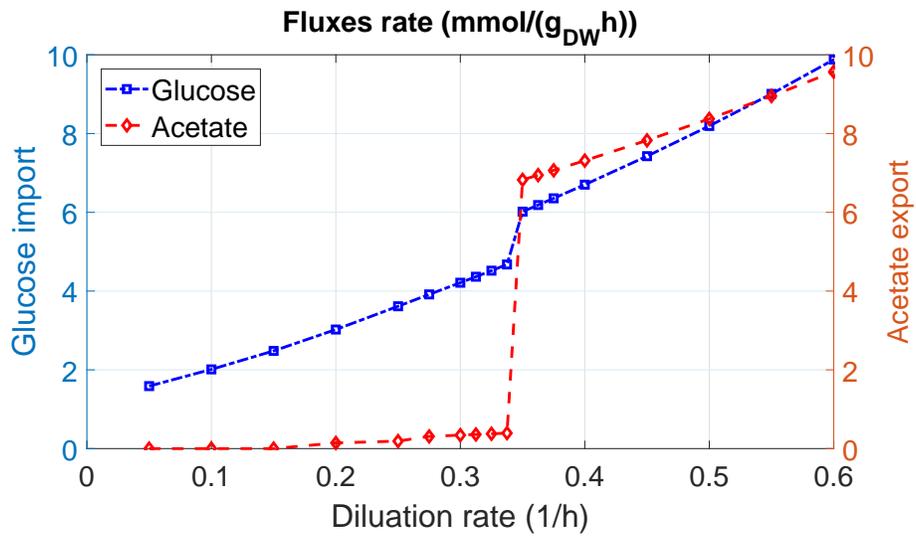}
\caption{Import and export fluxes in glucose-limited conditions}
\label{fig:glc_limited_flux}
\end{figure}

We have furthermore depicted on Fig.~\ref{fig:glc_limited_flux} 
the evolution of the glucose import and acetate export fluxes as a function of the dilution rate. The increasing nature of the glucose import and the existence of a switch on the acetate production is consistent with experimental data presented in \cite{DaS:01,DSS:01} even if the predicted switch appears for a dilution rate between 0.3 and 0.4 $h^{-1}$ but experimentally between 0.2 and 0.3 $h^{-1}$ (see Fig. 5D in  \cite{DaS:01}). This switch highly depends on the respiratory system whose parameter calibration may lack of precision due to the lack of data on the membraneous proteins.\ 
\begin{figure}[thpb]
\centering
\includegraphics[width=12cm]{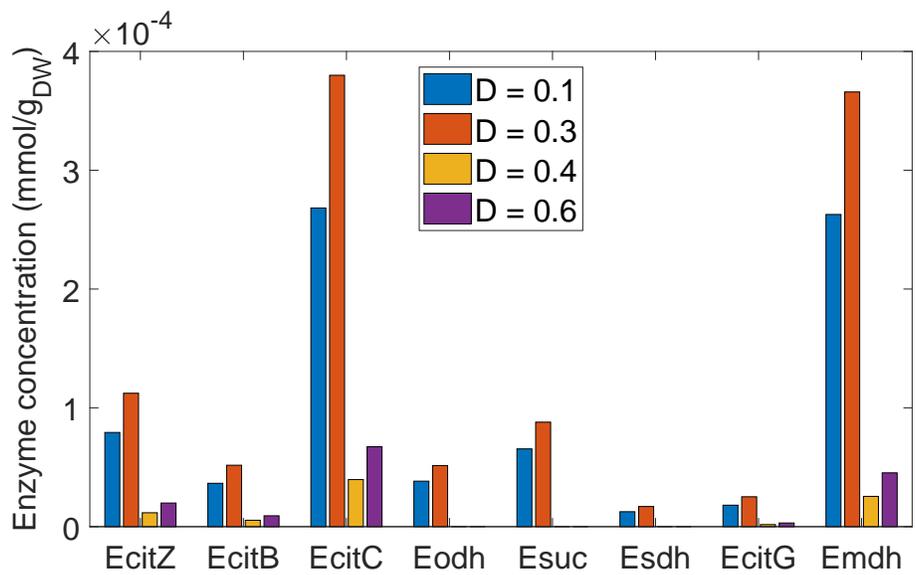}
\caption{TCA enzymes concentration in glucose-limited conditions}
\label{fig:glc_limited_TCA_invest}
\end{figure}

More generally, since we  have access to the internal states of the bacterial cell, it is possible to observe on the prediction that the switch for acetate production is associated to a modification of the TCA cycle enzyme concentrations (see Fig. \ref{fig:glc_limited_TCA_invest}). This last observation is in accordance with experimental data and with the status of the catabolite repression in chemostat (see \textit{e.g.} \cite{DSS:01} and \cite{DeF:94}).   
\begin{figure}[thpb]
\centering
\includegraphics[width=12cm]{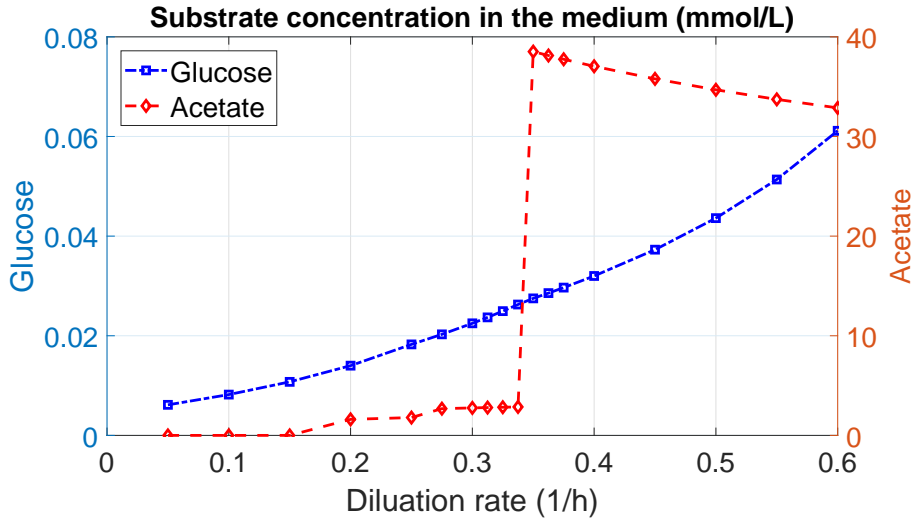}
\caption{Glucose and acetate concentration in culture medium in glucose-limited conditions}
\label{fig:glc_limited_medium}
\end{figure}
It is expected that the concentration of the  glucose in the culture medium is increasing as a function of the dilution rate (see Fig. \ref{fig:glc_limited_medium}). Indeed, since the glucose is the limiting substrate and since the transporter efficiency is a Michaelis-Menten function, a higher transporter efficiency is obtained by a higher glucose concentration in the culture medium. The acetate concentration confirms the computed export rate, it is low below 0.3 $h^{-1}$ of dilution rate whereas it is high above 0.4 $h^{-1}$. It is surprising however that the acetate concentration decreases above a dilution  0.4 $h^{-1}$: actually the increase in export does not compensate for the higher dilution rate.
\begin{figure}[thpb]
\centering
\includegraphics[width=12cm]{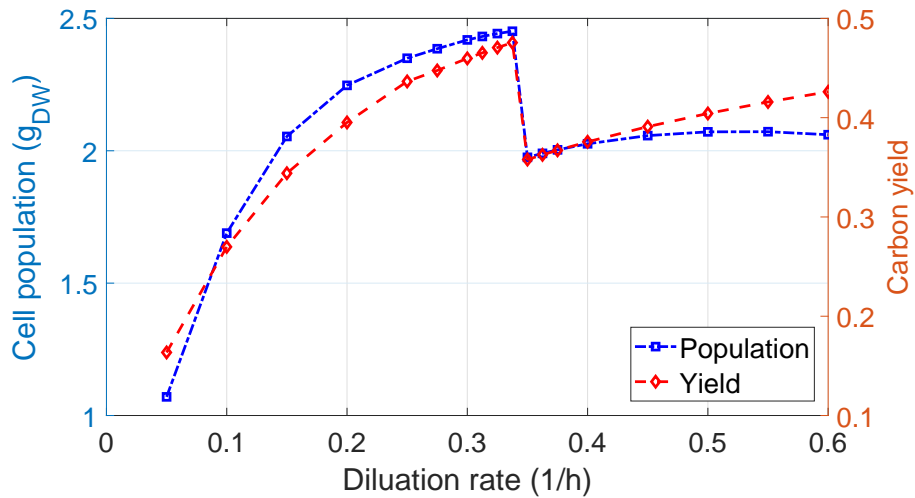}
\caption{Cell population  and carbon yield in glucose-limited conditions}
\label{fig:glc_limited_population}
\end{figure}

Fig. \ref{fig:glc_limited_population} displays the cell population. Here again, there is a change in behavior with a decrease of population between 0.3 and 0.4. This decrease is related to the export of acetate and the change of the biomass production yield. This change cannot be predicted by Monod-type models. Indeed, a part of the  carbon flux, imported through the glucose in the cell, is exported via acetate instead of being used for the biomass or energy production: the yield is necessarily lower when acetate is produced.

In fact, when the dilution rate changes, the distribution of carbon between biomass, respiration and overflow dramatically changes as shown in Fig.~\ref{fig:glc_limited_end_product}. We stress the importance of maintenance cost at a low dilution rate in the distribution of carbon. Indeed the energy requirements for cell maintenance lead to an increase in the proportion of glucose used to produce energy at the expense of biomass when the dilution rate is decreasing.

\begin{figure}[thpb]
\centering
\includegraphics[scale=0.1222, trim = 0 0 0 20, clip]{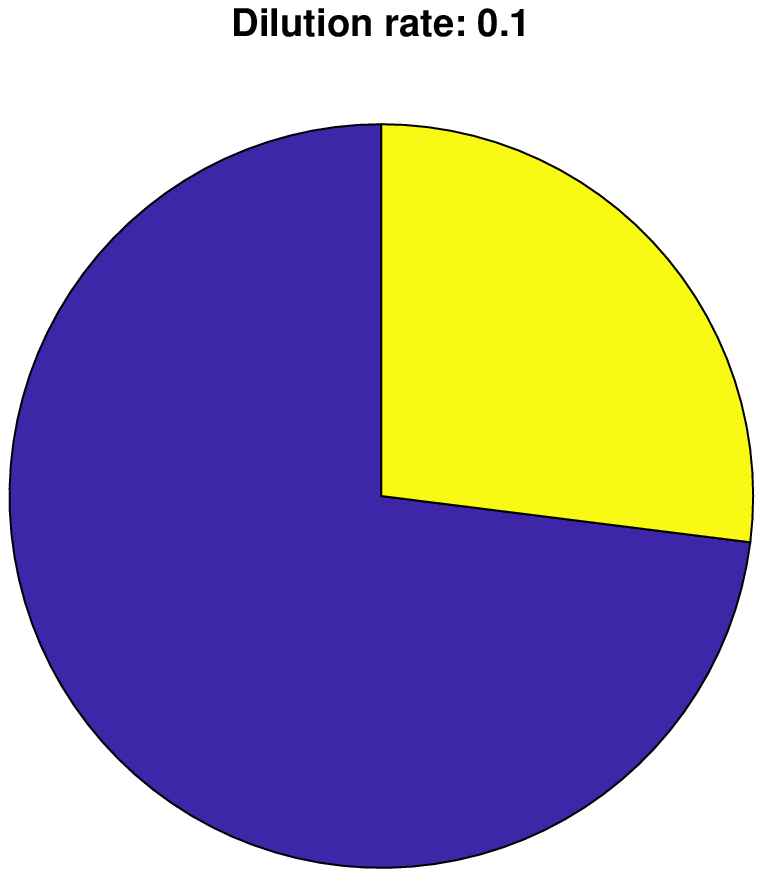}
\includegraphics[scale=0.2558, trim = 0 0 0 10, clip]{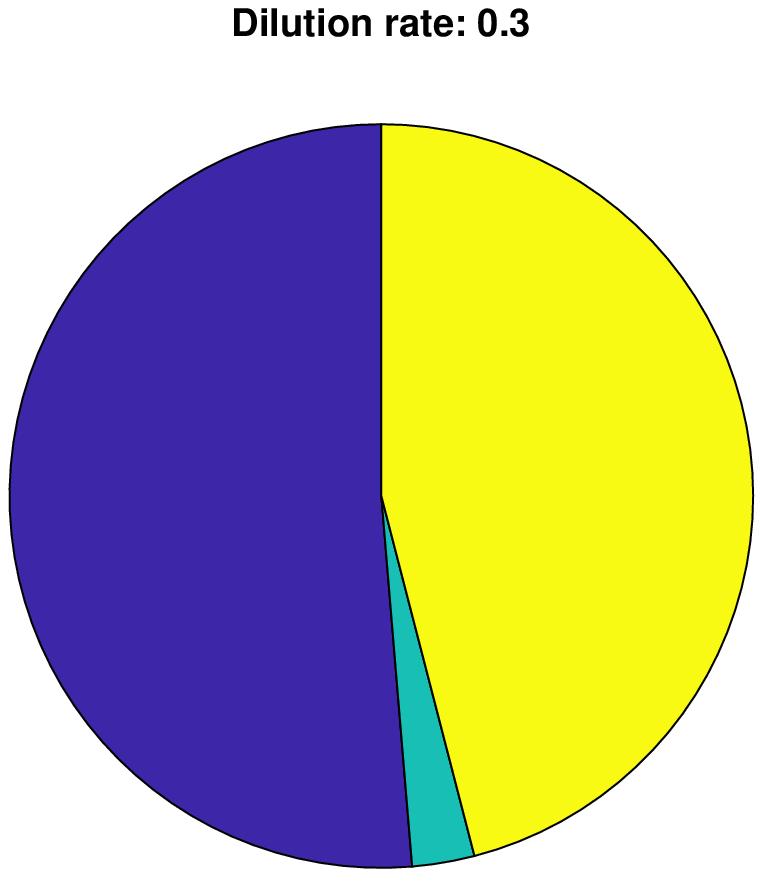}
\includegraphics[scale=0.4072, trim = 0 0 0 10, clip]{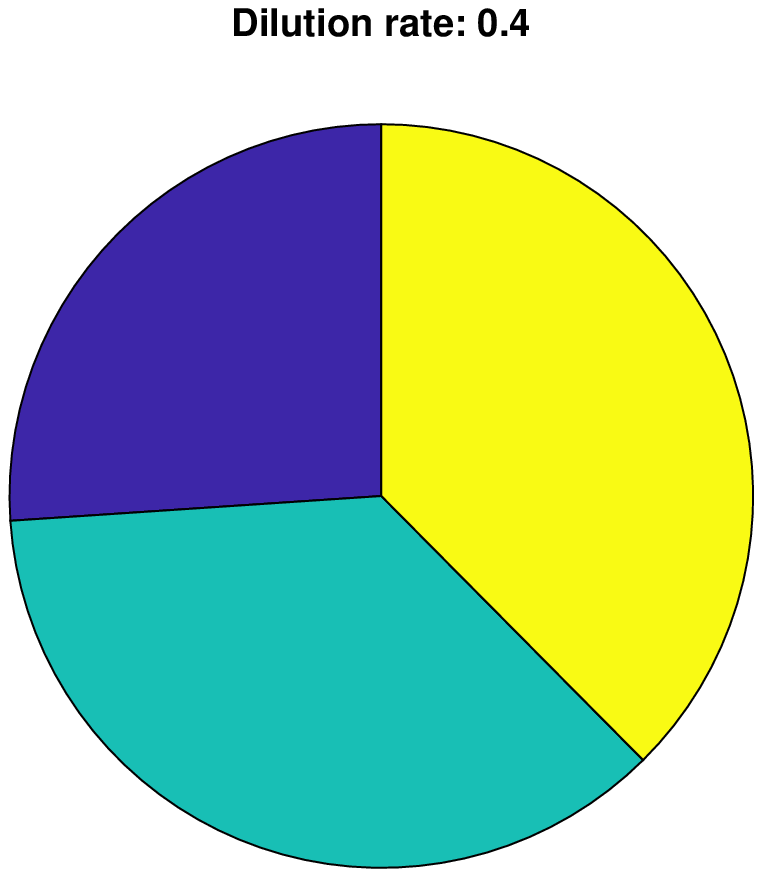}
\includegraphics[scale=0.6, trim = 0 0 0 10, clip]{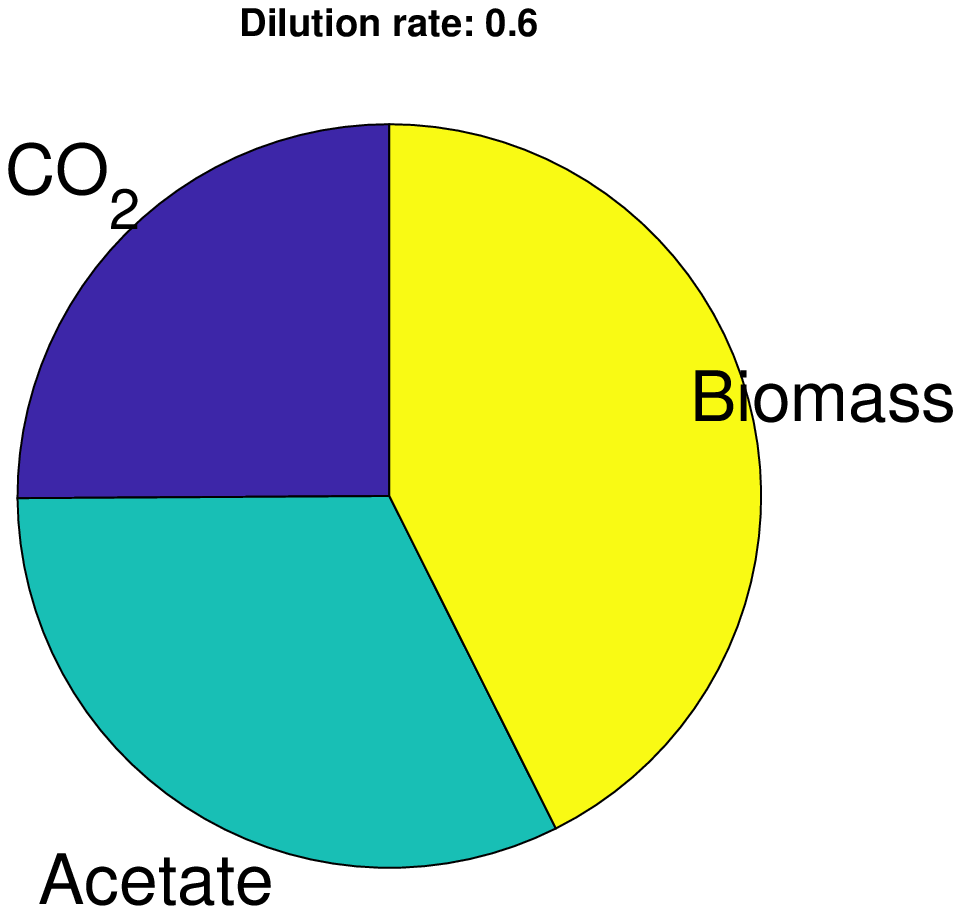}
\caption{Glucose-limited conditions: carbon end-product (from left to right, dilution rate: 0.1, 0.3, 0.4 and 0.6 $h^{-1}$). The circle radius is proportional to the uptake of glucose.}
\label{fig:glc_limited_end_product}
\end{figure}

\subsection{Chemostat in sulfate-limited conditions}
\begin{figure}[thpb]
\centering
\includegraphics[width=12cm]{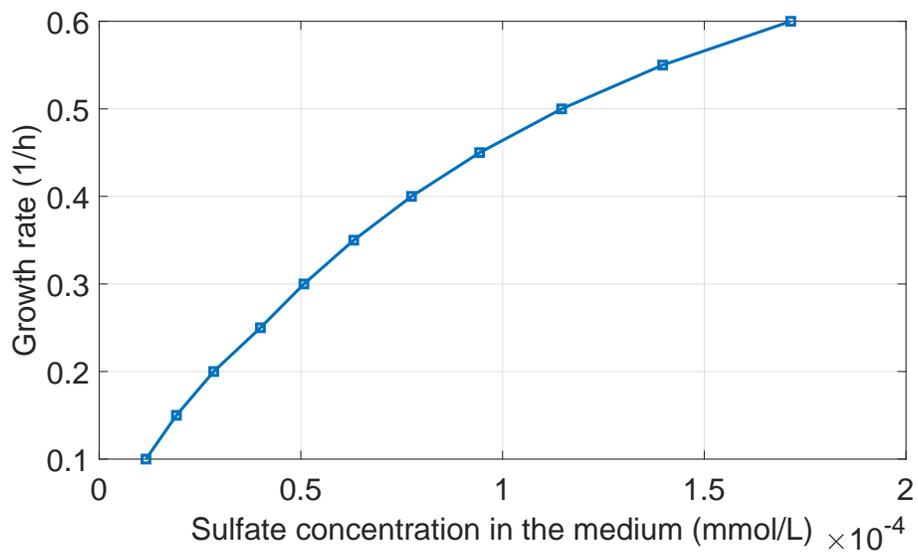}
\caption{``Monod's curve'' in sulfate-limited conditions}
\label{fig:sul_limited_monod}
\end{figure}

\begin{figure}[thpb]
\centering
\includegraphics[width=12cm]{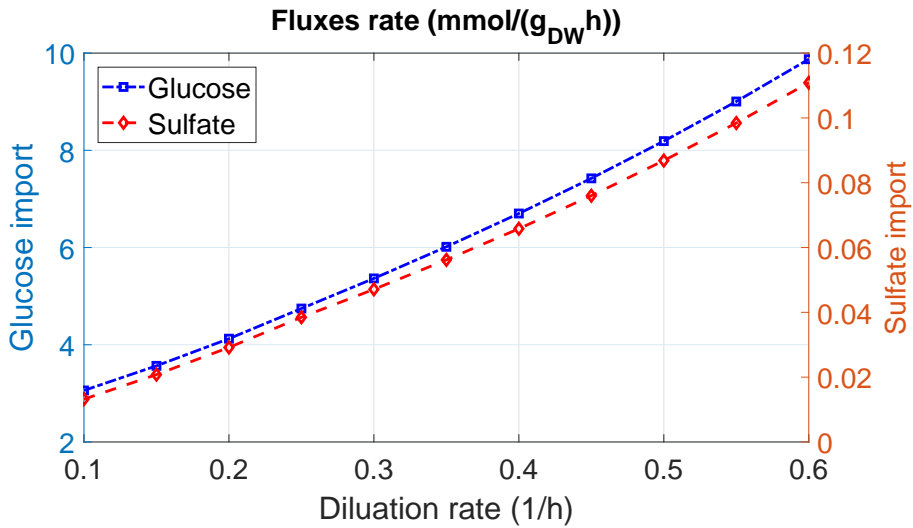}
\caption{Glucose and sulfate import fluxes in sulfate-limited conditions}
\label{fig:sul_limited_flux}
\end{figure}

As seen in the previous section, the prediction of bacterial cell behavior towards a limited--carbon source leads to complex behaviors due to the dual role of the carbon source in growth, i.e. as a precursor of biomass but also as a source of cell energy.  In order to obtain a less complex behavior, it is sufficient to take as limiting source a substrate that has a role only in the formation of biomass. We consider that the inflow medium is the same than in the previous section except that the sulfate is now the limiting nutrient. To this end, its initial and non--limiting concentration was reduced from 47 to a limiting concentration around 0.1 $mmol/L$.\\
As expected, the behaviour of the bacterial cell as a function of the dilution rate is much more regular in this case.  Fig. \ref{fig:sul_limited_monod} shows once again Monod's curve related to sulphate--limited conditions. As shown in Fig. \ref{fig:sul_limited_flux}, the fluxes of glucose and sulphate are both increasing almost linearly as a function of the dilution rate of the chemostat. The slight nonlinearity is due to the modification of the bacterial cell composition through the increase in the number of ribosomes needed for supporting ``higher'' growth rate.  
\vspace*{-0.3cm}

\subsection{Switch of limiting substrate}
\begin{figure}[thpb]
\centering
\includegraphics[width=12cm]{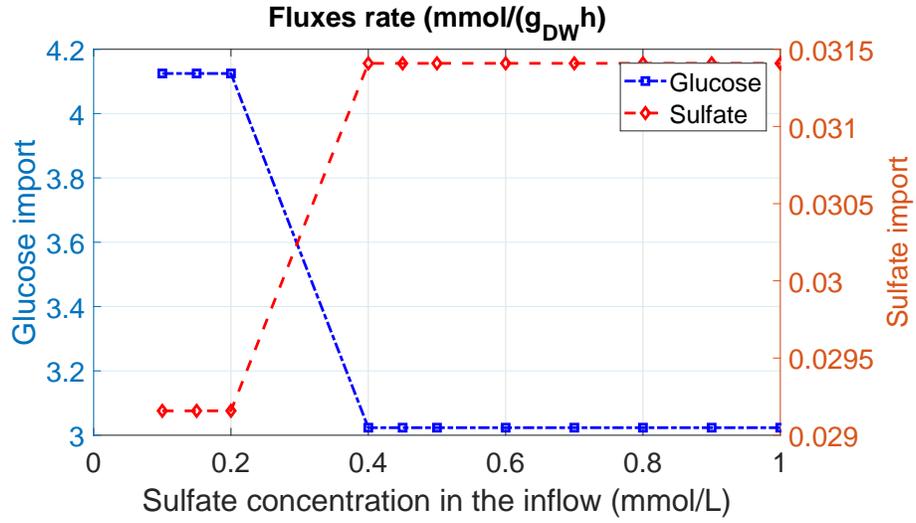}
\caption{Glucose and sulfate import fluxes for varying inflow composition}
\label{fig:sul_D_limited_flux}
\end{figure}

\begin{figure}[thpb]
\centering
\includegraphics[width=12cm]{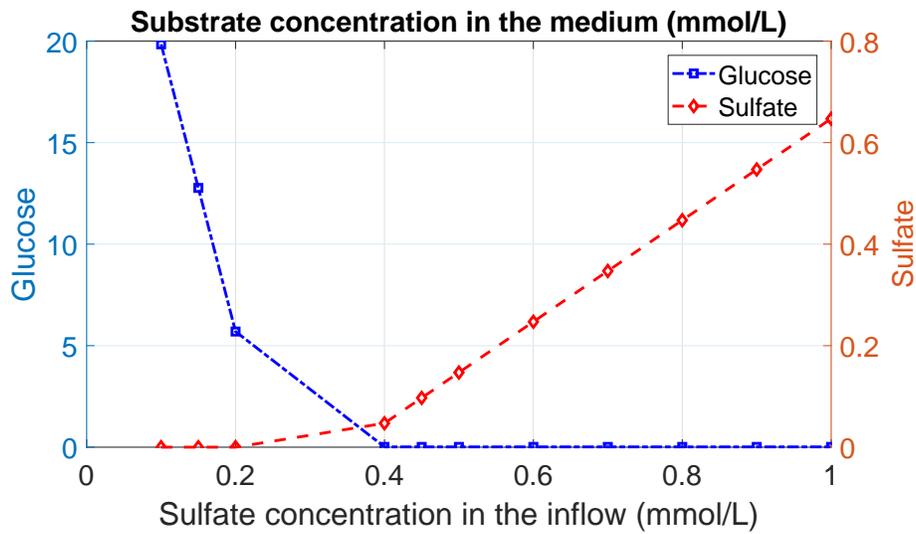}
\caption{Glucose and sulfate concentrations in the medium for varying inflow composition}
\label{fig:sul_D_limited_medium}
\end{figure}

These illustrations of the interest of using the RBA approach for the chemostat are concluded by considering a scenario where the limiting nutrient in the inflow changes. 
This experience is classic and consists in changing the inflow composition at a given dilution rate. The results illustrate that the RBA approach recovers the behavior of the chemostat  for a so--called dual limitation, defined between carbon and nitrogen substrates in \cite{Egl:91}, here between glucose and sulfate (see Fig. \ref{fig:sul_D_limited_flux} and Fig. \ref{fig:sul_D_limited_medium}).

\medskip

\section{Conclusion} \label{sec:conclusion}
In this article, we have proposed a new chemostat model by replacing the phenomenological and classic Monod's model (and the variants) by an RBA type model.  As in the case of the RBA method in batch mode, we have proved in this article that the chemostat problem is also a convex problem under realistic assumptions. 
This implies that the problem which involves a few thousands of decision variables can be easily handled (as it is the case here, where our RBA model contains more than one thousand decision variables). 
Finally, the first tests of this new method show surprising matches with experimental observations in qualitative terms for a wild-type strain; these matches are impossible to obtain with Monod' type model because of its intrinsic simplicity. In addition, it has been demonstrated in \cite{GMC:15} that the RBA method also has predictive capabilities for modified strains so that the proposed model can certainly and advantageously be used in a  biotechnological context.

\section{Acknowledgement}
Finally, we thank Anne Goelzer for her work on the  RBA model and providing us with it. We also thank the reviewers for their useful comments.

\printbibliography

\section{Appendix}
\subsection{Brief description of the RBA model constraints} \label{sec:rbaDesc}
At a constant growth rate $\mu$, the cell population evolves as $X(t) \eqbydef X(0)e^{\mu t}$ and all the concentrations in the cells are constant. In a first approach, a cell can be viewed as being composed of two parts: $i)$\ the metabolic network, comprised of $E$ and $T$ and the corresponding fluxes $\eta$ and $\nu$, and $ii)$ the translation apparatus in a broad meaning, comprised of $R$ and $C$. To keep the concentration of proteins constant at a growth rate $\mu$, the cell needs to produce the proteins at a rate equal to $\mu E_i$, $\mu T_i$, $\mu R$ and $\mu C$. With this in mind, we now briefly describe the RBA problem in batch mode.

Even if this is not directly apparent, the constraints $(C_1)$ are equality of fluxes (per unit of cell population) and correspond to the conservation of mass of internal -- to the cell population -- metabolites
: any metabolite produced is consumed or exported in the medium, hence the equality to $0$. In order to facilitate the understanding, these metabolites are classified into 4 groups denoted by $p$ for precursor, $c$ for constant concentrations, $r$ for recycled and $i$ for the other internal metabolites. The precursor metabolites are used by the translation apparatus while the recycled metabolites are produced by it, hence the difference of sign between $a)$ and $c)$. It should be noted that the precursor and recycled metabolites are at the interface between the metabolism and the translation apparatus. In this context, the matrices $C_\star^{p/r}$ correspond to the composition in metabolites of the corresponding entity, with the scalar $P_G\geq 0$ corresponding to a set of proteins linked to stress responses or other functions not included in the current model. The constraints $a)$ and $c)$ thus define an equality of fluxes -- corresponding to the necessary ones to keep the protein concentrations constant -- at the interface between the metabolic network and the translation apparatus. The metabolites in the group $c$  typically correspond to macromolecules such as the lipid and peptidoglycan necessary to the membrane and wall synthesis. In this context, $\mu\bar{X}^c$ corresponds to the necessary flux for the synthesis.

$k_T$ and $k_C$ being the production efficiency of the ribosome $R$ and the folding efficiency of the chaperon $C$, the constraints $(C_2)$ corresponds to the fact that $R$ (and $C$) are high enough to produce (and fold with $\alpha_C$ being the percentage of proteins that actually need the help of the chaperon to fold) the desired quantity of proteins, equal to the $\mu E$ for the enzymes for instance. In this context, the matrices $C_\star^\star$ correspond to production cost.

$D^D$ and $D^S$ being saturating occupancies, the constraints $(C_3)$ correspond to the fact that the quantity of proteins is limited (the vectors $C_\star^\star$ correspond then to the size of the entities) by the available space, the volume of the metabolites being neglected.

The constraints $(C_4)$ correspond to the limitation of the reaction flux by the amount of enzyme that catalyzes the reaction (with $\underline{k}_i$ and $\overline{k}_i$ being the efficiencies in the backward and forward directions of the enzyme). The constraint $(C_5)$ is a finer description of constraints $(C_4)$ for the transporters.

\subsection{The efficiency of the transporters is parameterized by a Michaelis-Menten} \label{sec:discuss TransporterEfficiency}
We explain here why it is possible to use simple models for
modelling the transporter efficiency in spite of the existence
of different types of passive and active transporters in the cell.

\paragraph{The second law of chemical thermodynamics.} Like all biochemical reactions, the transport of a medium substrate across the cell membrane  must satisfy thermodynamic constraints. For a 
general biochemical reaction defined by 
$$
(A)+(B) \rightleftharpoons (C) + (D)
$$
where $(A)$ ,$(B)$, $(C)$ and $(D)$ are given molecules\footnote{The usual notations are $X$ for the molecule and $[X]$ for its concentration. To be consistent, we do not follow this usual notation.}, the Gibbs free energy associated to this reaction is defined by  
\begin{equation}
\Delta G =  \Delta G^0  +  RT \ln \frac{CD}{AB}  \label{thermodynamic_constraint}
\end{equation}
with $X$ denoting the concentration of the molecule $(X)$ and where $\Delta G^0$ corresponds to the standard free energy of the reaction ($J$)\footnote{The units are: $J$ for joule and $K$ for kelvin.}.
$\Delta G^0$ is related to the equilibrium constant $K_{eq}$ of the reaction (unitless) by the relation $\Delta G^0 =-RT \ln K_{eq}$ with $R$ the ideal gas constant ($J/mol/K$) and $T$ the temperature ($K$). In order for the reaction to have a net flux of $C$ and $D$ production in a steady-state regimen, i.e. the net flux is in the forward direction, it is necessary that $\Delta G < 0$.

\paragraph{Passive transporters.} The reaction associated to an uniporter $T$ is a passive (non-energetic) transport given by 
$$
(S)_{m}  \stackrel{T}{\rightleftharpoons} (S)_{c}
$$
where $(S)_m$ and $(S)_c$ denote respectively the molecule $(S)$ in the medium and in the cell.
Since the molecule in both sides are the same, the standard free energy of the reaction is $0$, \textit{i.e.} $\Delta G^0 = 0$. A net import flux of $(S)$ inside the cell is possible only if
$$
\ln \frac{S_c}{S_m} < 0
$$
that is when the concentration in the medium is higher than the concentration in the cytosol: $S_c < S_m$. It is worth noting  that in the inverse case, the flux through the transporter is reversed leading to export the molecule in the medium. That explains why it   is often suggested in the literature that the molecule associated to a given uniporter is generally immediately modified inside the cell in order that its  concentration in the cytosol remains close to $0$. This mechanism then traps the modified molecule within the cell. 
A classic example is given by the glucose uniporter of yeast and the immediate phosphorylation of the imported glucose by a glucokinase into 
glucose 6--phosphate, which cannot freely cross the membrane as a phosphorylated molecule. 

Following a classical way to describe enzyme kinetics, the efficiency of an uniporter is well described by a Michaelis-Menten type function defined by
$$
k(S_m) \eqbydef \frac{VS_m}{K+S_m}
$$
which writes with our notation as
\begin{equation} \label{eqn:michaelisMenten}
 k(S) \eqbydef \dfrac{V S }{K+S}
\end{equation}
where $V\geq 0$ is  the maximal activity of the transporter and $K\geq 0$ is the  Michaelis-Menten constant of the transporter.

\paragraph{Active transporters powered by the proton--motive force.} The uniporters are not the most common transporters in bacteria since a large part of bacterial transporters are active (energized). 
Among them, the symporters coupling the import of a  substrate of interest with the import of a proton ($\text{H}^+$) are the most common. The energy associated to this class of transporters is related to the so--called proton--motive force (pmf) and the general theory of chemiosmosis \cite{Mit:61}.  
Actually, when a gradient of electrical potential exists between the two sides of the membrane, the Gibbs free energy of the reaction takes a different form. For example, when only a proton is imported, i.e.
$$
(\text{H}^+)_{m}  \stackrel{T}{\rightleftharpoons} (\text{H}^+)_{c} ,
$$ 
the Gibbs free energy is given by  
\begin{equation}
\Delta G \eqbydef  F \Delta \psi +  RT \ln \frac{\text{H}^+_c}{\text{H}^+_m}  \label{thermodynamic_constraint_pmf}
\end{equation}
where $\Delta \psi$ corresponds to the difference in the electrical potential between the two sides of the membrane (by convention 
$\Delta \psi >0$ when the electrical potential in the medium side is lower than in the cytosol side) and $F$ is the Faraday constant. The concentration of proton has a specific status  in chemistry, leading to introduce the pH notion: $\text{pH} \eqbydef -\log_{10}\text{H}^+$. We can then rewrite \eqref{thermodynamic_constraint_pmf} as
\begin{equation*}
\Delta G = F \Delta \psi -  RT \, \ln(10)  \Delta \text{pH}  \label{thermodynamic_constraint_pmf_ph}
\end{equation*}
where  $\Delta \text{pH} \eqbydef  \text{pH}_{c}-\text{pH}_{m}$. The pmf is defined by  
$$
\Delta p \eqbydef -\Delta \psi  + \frac{RT \, \ln(10)}{F}\Delta \text{pH}.
$$
Obviously, an important actor in the definition of the pmf is the balance between the flux of proton export by the respiration chain and the flux of import mainly by the ATPase, but also by active transports to cite a few.  It should be noted that the import of protons linked to these active transports would lead without an active control to an acidification of the bacterial cytosol. The pmf is under a sophisticated and active control regulating its level despite the possible variations of the pH of the medium or the osmotic gradient between the medium and the cell.

When the import of a specific substrate is coupled with the one of a proton, \textit{i.e.}  
$$
(S)_{m} + (\text{H}^+)_{m}    \stackrel{T}{\rightleftharpoons} (S)_{c} +  (\text{H}^+)_{c},
$$
the Gibbs free energy is given  by
\begin{equation}
\Delta G =   -F \Delta p + RT \ln \frac{S_{c~}}{S_{m}}. \label{thermodynamic_constraint_pmf_substrate}
\end{equation}
By assuming that the pmf is regulated,  it is then not restrictive to assume in a first approach that the efficiency of such a transporter is also  well described by a Michaelis-Menten function of the substrate where the pmf is incorporated in the transporter efficiency as a multiplicative constant effect.

For a symporter coupling the import of a substrate with a sodium cation ($\text{Na}^+$) instead of a proton, the same approach can be used leading to define 
\begin{equation} \label{thermodynamic_constraint_pmf_Na_plus}
\Delta G =   F \Delta \psi +  RT \ln \frac{\text{Na}^+_{c~}}{\text{Na}^+_{m}} + RT \ln \frac{S_{c}}{S_{m}} .
\end{equation}
Actually, $\text{Na}^+$ is generally a toxic cation when present in the bacterial cell, and thus there exists a dedicated mechanism (an antiporter coupling the export of  $\text{Na}^+$ with an import of protons) exporting $\text{Na}^+$ from the cell into the medium leading to keep the concentration of $\text{Na}^+$ in the cytosol very low in a steady-state regimen. Like other active transporters,  the  efficiency of this transporter can be described by a Michaelis-Menten function of the substrate of interest by incorporating the pmf as a multiplicative constant effect in the efficiency of the transporter.

\paragraph{Active transporters powered by ATP hydrolysis.} 
Another important category of active transporters are the ones using the hydrolysis of ATP as an energy source. 
They couple the import of a specific substrate $S$ to the hydrolysis of ATP\ into ADP and Pi where Pi is an inorganic phosphate, \textit{i.e.}  
$$
(S)_{m} +  (\text{ATP})_{c} + (\text{H}_2\text{O})_c \stackrel{T}{\rightleftharpoons} (S)_{c} +  (\text{ADP})_{c} + (\text{Pi})_{c}
$$
The  Gibbs free energy is then given by\footnote{Note that the water, $\text{H}_2\text{O}$, is not used in the definition of the Gibbs free energy since it is the solvent.}    
\begin{equation} \label{thermodynamic_constraint_ABC}
\Delta G =  \Delta G^0  +   RT \ln \frac{\text{ADP}_{c} \text{Pi}_{c}}{\text{ATP}_{c}} + RT \ln \frac{S_{c}}{S_{m}}
\end{equation}
where $\Delta G^0$ corresponds to the standard free energy of ATP hydrolysis into ADP and Pi, i.e. $\text{ATP} + \text{H}_2\text{O} \rightleftharpoons \text{ADP} + \text{Pi}$, with a value close to $35-40$ $kJ/mol$.

Like in the previous cases, the cell tightly regulated the energy level in the cell, leading in a first approach to assume that the efficient of such a transporter is also well described by Michaelis-Menten like function where the energy level of the cell only implicitly appears as a constant effect.

\subsection{Each substrate is either imported or excreted but not both at the same time} \label{sec:one_transporter}
We discuss hereafter the assumption that the bacterial cell
does not use transporters at the same time to import and
export the same substrate. This hypothesis is generally satisfied by a bacterial cell in a steady-state regimen (the case of
proton associated to transporter is discussed below). Indeed,
when this assumption is not satisfied, it leads generally to
the definition of a so-called futile cycle where energy is
dissipated for nothing. Indeed, in order that two opposite
transports to be possible, at least one of the transporters has
to use energy for thermodynamical reasons.
However, there exists a situation where such a futile cycle
could appear: when a molecule can diffuse freely through
the cell membrane (two classic examples are the acetate and
the $\text{NH}_4^+$). Here without going into too much details, it can
be proved that this case leads to a convex formalization
of the chemostat problem (since diffusion is free and do not
necessitate a transporter).
It mainly remains to consider the exception related to the
protons. Indeed, in many conditions, the protons are massively exported into the environment by the respiratory chain
and also massively re-imported into the cell by the ATPase
in order to produce energy (ATP) or, as we saw in the
previous section, used as a way to energize the import of
some substrates. So formally the proton does not satisfy our
main assumption. Actually, that does not lead to a major
problem since in most of the chemostat experiments, it is
necessary to tightly control the pH of the
culture medium by external means. This means that the concentration of proton
in the culture medium is no longer a decision variable but a
given constant.

\end{document}